\documentstyle[11pt,fullpage,doublespace,epsf]{article}
 \DeclareMathSizes{11}{19}{13}{9}   

\makeatletter
\@addtoreset{equation}{section}
\makeatother

\begin{document}

\title{Computational Exploration\\
 of the\\
 Nanogold Energy Landscape}

\author{Michael McGuigan\\
and\\
J.W. Davenport\\
Brookhaven National Laboratory\\Upton NY 11973\\mcguigan@bnl.gov}
\date{}
\maketitle

\begin{abstract}
We use density functional theory to quantify finite size and shape effects for gold nanoclusters. We concentrate on the computation of binding energy as a function of bond length for icosahedral and cuboctohedral clusters. We find that the cuboctoheral gold clusters have lower energy for 13 atoms. For 55 atoms we find that the icosahedral gold clusters have lower binding energy. We also introduce a one parameter family of geometries that interpolate between the icosahedral and cuboctohedral clusters that is parametrized by an angle variable. We determine the binding energy dependence on shape as a function of the  angle variable  for 13 and 55 atom clusters with a minimum at the cuboctohedral point and icosahedral point respectively. We also compute the binding energy for the 147 atom gold nanocluster and show that the binding energy of the icosahedral cluster is lower than the 147 atom cuboctohedral gold cluster. We also compute the binding energy of the $Au_{55}O_2$ molecule with possible applications to catalysis.
\end{abstract}

\section{Introduction}

An important application area for nanoscience are gold nanoclusters. These clusters can be used in energy science as catalysts and in medicine for drug delivery. It is also important to understand the basic physics and chemistry of gold nanoclusters to understand the limitations of the current applications as well as to suggest new ones. Experiments on gold nanoclusters have recently been performed on clusters with 102 gold atoms \cite{Jadzinsky}. In addition increasing computational power has lead to the ability to simulate a nanocluster of hundreds of atoms on supercomputers running scalable density functional codes. Thus this is an exciting time in the development of nanoscience where computation, theory, and experiment are helping to improve our understanding of these systems.

The main tool that we use to computationally probe gold nanoclusters is density functional theory run on large parallel computers. Density functional theory is a computationally efficient approach to many body systems that maps the interacting many body Schrodinger equation to a single body noninteracting problem given by:
\[
\left(-\frac{\hbar^2}{2m} \bigtriangledown^2 + V_s(\overrightarrow{r})\right)\psi_i(\overrightarrow{r}) = \epsilon_i \psi_i(\overrightarrow{r})
\]
where $V_s$ is defined by:
\[
V_s(\overrightarrow{r}) = V_{ext}(\overrightarrow{r}) + \int \frac{e^2 n(\overrightarrow{r})}{|r-r'|} d^3r' + V_{XC}[n(\overrightarrow{r})]
\]
with $V_{ext}(\overrightarrow{r})$ the external potential, $V_{XC}[n(\overrightarrow{r})]$ the exchange correlation potential and $n(\overrightarrow{r})$ the electron density.

One typically uses a local density approximation and generalized gradient approximation where the exchange functional is a local function of the density and gradient of the density. One then iteratively solves the one-body equation using the relation between the electron wave function and density to evaluate the the exchange correlation potential. More details about the density functional approach can be found in many textbooks, for example \cite{Fiolhais}.

Density functional theory has been implemented in several quantum chemistry codes. In this paper we use the NWChem quantum chemistry code distributed by PNNL \cite{nwchem}. We also use a large Blue Gene L parallel supercomputer that efficiently runs NWChem up to 2048 nodes or 4096 processors. A general description of the application of NWChem to materials science can be found in \cite{Apra1}. Outstanding features of the NWChem input file specify the geometry, basis, effective core potential, exchange and correlation functionals, DIRECT, SMEAR and MULT. We used the basis set specified by SBKJC VDZ ECP \cite{Stevens} and the pbe exhange and correlation functionals \cite{Perdew}. The DIRECT keyword indicates that all integrals are computed on the fly, the SMEAR keyword allows fractional occupation of molecular orbitals, and the MULT keyword defines the spin multiplicity \cite{nwchemuser}. In this paper we used the default smear value of .001 and multiplicity 2. 

Besides the practical application of gold nanoclusters little is known of the structure of the energy landscape of theses systems even though the chemistry and material properties strongly depend on the position in this landscape. The main difficulty is that even with 13 atoms one has a $13 \times 3$ dimensional space to explore and this is difficult computationally as well as difficult to visualize. Thus the main purpose of this paper is to find a simpler description of this landscape in terms of the size and shape of the gold nanoparticles for well known structures like the icosahedron and cuboctohedron. Previous density functional studies of gold nanoparticles are given by \cite{Haberlen}, \cite{Wang}, \cite{Baletto},
\cite{Dong}, \cite{Gruber}. In particular \cite{Haberlen} and \cite{Wang} studied the relative energy between icosahedral and cuboctahedral gold clusters and our results are consistent with their results. The nanogold phase map in temperature and critical size has recently been determined based on relativistic ab initio calculation and verified using high resolution electron microscopy at finite temperature  \cite{Barnard}. Their work indicates that different shapes of nanoparticle are more stable depending on the place in the phase diagram. By  varying the size and shape of the gold naocluster we are able to determine the stable configurations as a function of critical size using density functional methods at zero temperature.

For gold, relativistic effects are very important. In \cite{Hakkinen} density functional studies indicate a qualitative difference between 55 atom gold and silver nanoclusters, and found gold nanoclusters have several low-lying low-symmetry structures and washed-out electron shell structure \cite{Hakkinen}. In \cite{Huang} photo-electron spectroscopy and first principle calculations  indicate a nonicosahedral disordered cluster with  strong surface contractions due to relativistic effects. In this paper we use an exchange potential that includes relativistic effects. However we focus on two well known structures cubocahedral and icosahedral. This leaves aside the question of the true lowest energy state of the system, but allows us to study the relative stability of the two structures as well as the transition and intermediate geometries between the icosahedron and cuboctahedron. These type of Mackay transitions between different shapes have been studied for lead \cite{Wei} and iron \cite{Rollmann} and we extend this analysis to gold nanoclusters here.

This paper is organized as follows. In section 2 we present the computation of the energy of 13 atom gold clusters in the shape of a cuboctahedron and icosahedron as well as intermediate structures. We derive a potentials which describes the dependence of the energy on the size and shape of the system. In section 3 we do the same thing for 55 atom gold clusters and compare with the previous section. We find that the cuboctahedron is more stable than the icosahedron for 13 atoms while the icosahedron is more stable than the cuboctahedron for 55 atoms. In section 4 we study the binding energy of the $Au_{55}O_2$ molecule and applications to oxidation and catalysis. We calculate the potential associated with the separation of the two oxygens and show that the presence of the gold nanoparticle lowers the potential relative to the free $O_2$. In section 5 we state the main conclusions of the paper.
 
\section{Finite size and shape effects in 13 atom gold nanoclusters}

In density functional theory (dft) one usually computes the energy $E$ of the multiple atom system and derives the binding energy $V$ from:
\[
V = E - N E_{atom}
\]
where $N$ is the the number of atoms and $E_{atom}$ is the dft computation of the energy of a single gold atom.

The total number of atoms in a cuboctohedron or icosahedron $N$ is given by \cite{Martin}:
\[
N = \frac{10}{3}K^3 - 5 K^2 + \frac{11}{3} K -1
\]
It Table 1 we list the first few values of $N$ for integer $K$ which is the shell index.

\begin{table}
\centering \caption{Number of atoms $N$ for a cuboctahedron or icosahedron for integer $K$ which is the shell index.}
\label{pdtable1}
\begin{tabular}{|c|c|}
\hline
K & N \\
\hline
2 & 13\\
3 & 55\\
4 & 147\\
5 & 309\\
\hline\end{tabular}
\end{table}

\begin{figure}[tbp]
   \centerline{\hbox{
   \epsfxsize=2.8in
   \epsffile{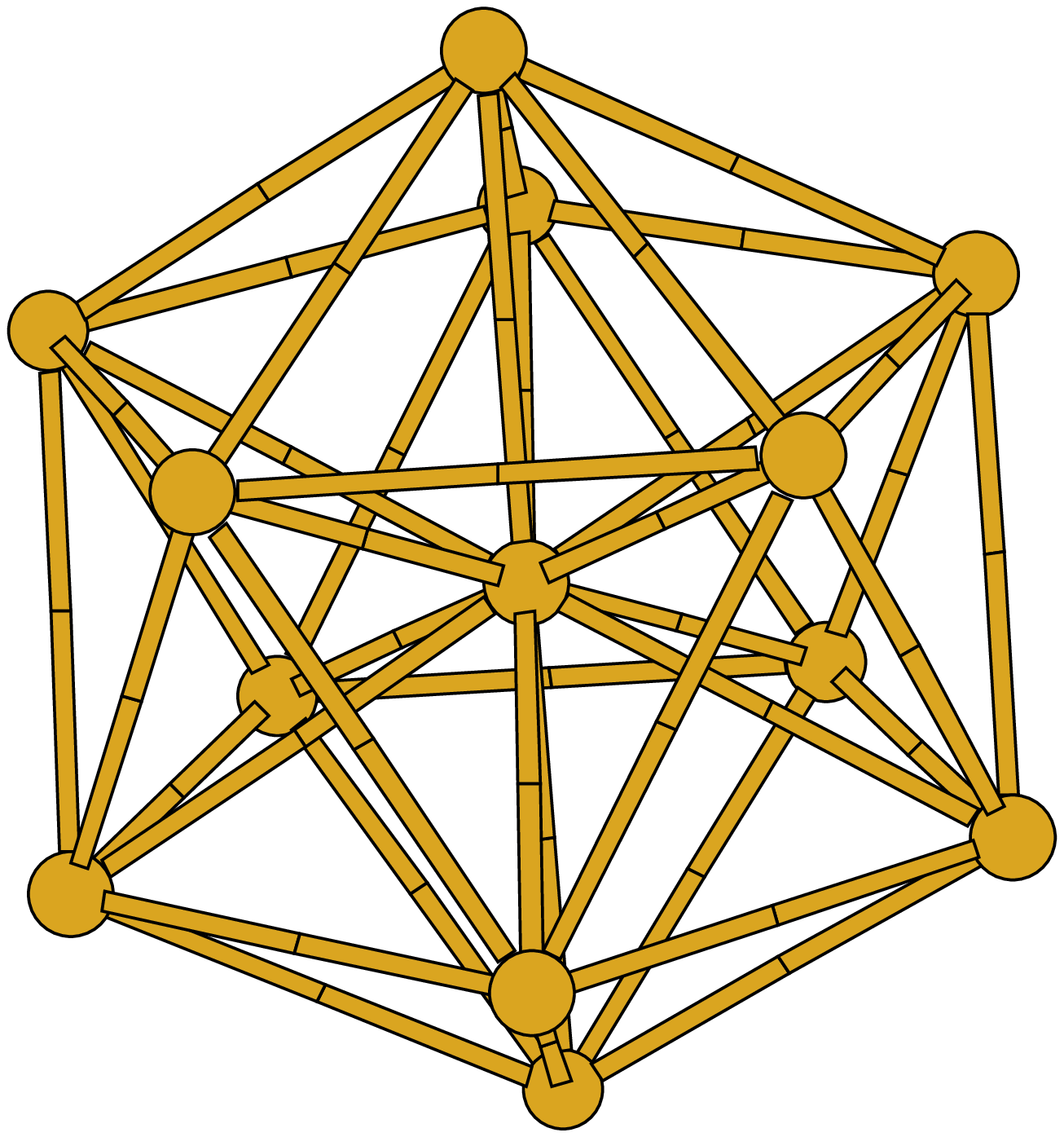}
   \epsfxsize=3.0in
   \epsffile{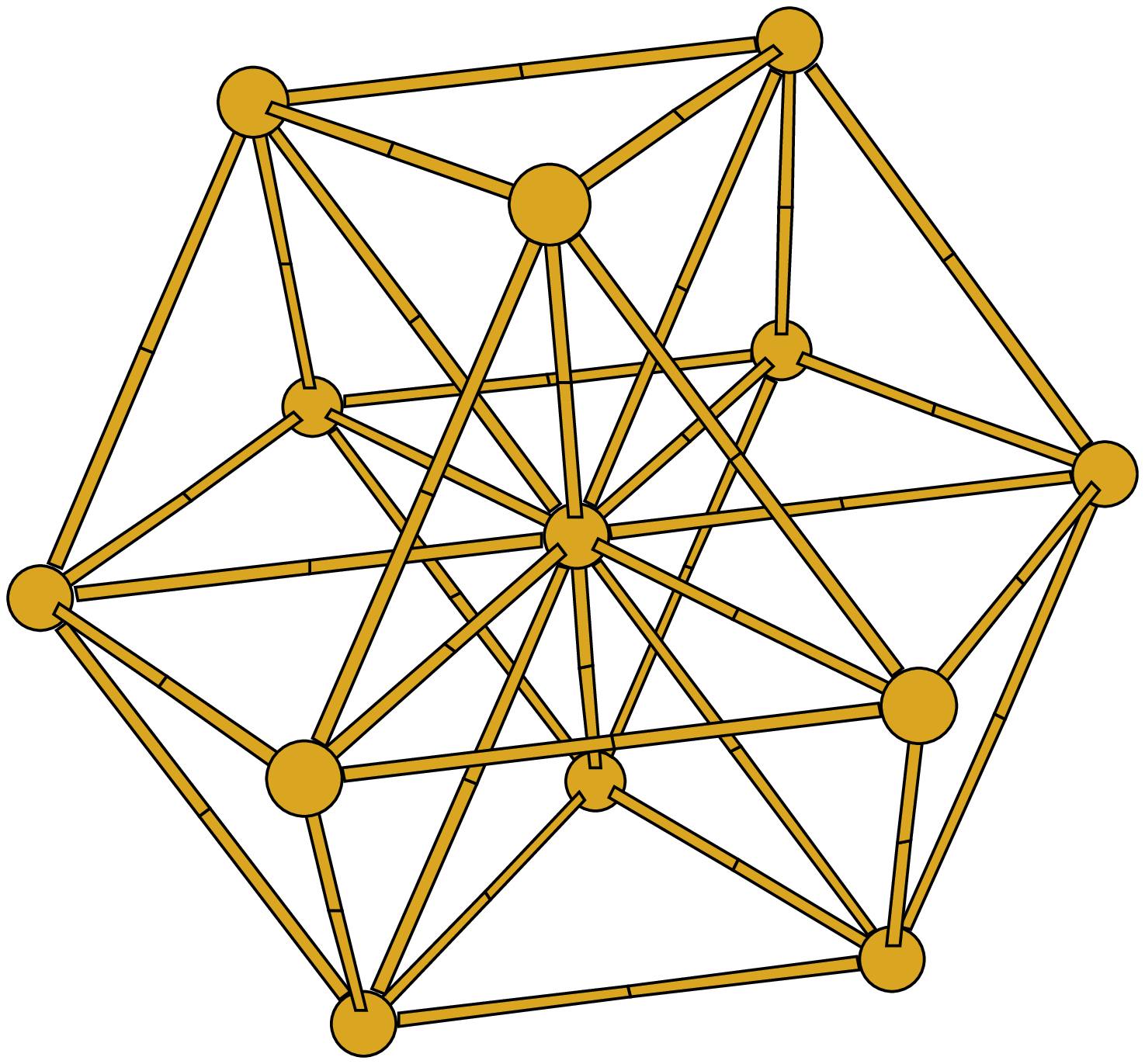}
     }
  }
 \caption{13 gold atom nanocluster in the shape of an icosahedron (left) and cuboctahedron (right).}

  \label{fig0}

\end{figure}

We show the 13 atom icosahedral and cuboctahedral geometry in Figure 1.
For the case of $K=2$ and $N=13$ we list the coordinates for the cuboctahedron in table 2  with bond length $a$. We list the coordinates for the 13 atom icosahedron also in table 2. Here $\varphi = \frac{1 + \sqrt 5}{2}$ is the golden ratio which is approximately $1.618033988749895$. For the icosahedron there are two bond lengths, $a$ and approximately $0.951057 a$ and we use the former. One can also consider intermediate shapes parametrized by a shape parameter $\theta$ whose coordinates are given in Table 2. Here $\theta = 45^{\circ}$ corresponds to the cuboctohedron and $\theta = \arctan \varphi \approx 58.282525589^{\circ}$  corresponds to the icosahedron. The list of coordinates for intermediate geometries is given in the third column of Table 2.

\begin{table}
\centering \caption{Coordinates for 13 gold nanoclusters with bond length $a$ and shape of cuboctahedron, icosahedron and intermediate structure with shape parameter $\theta$.}
\label{pdtable2}
\begin{tabular}{|c|c|c |}
\hline
cuboctohedron & icosahedron & intermediate \\
\hline
$a\frac{1}{\sqrt{2}}(0, 1, 1)$ & $a\frac{1}{{2}}(0, 1, \varphi)$ &  $a(0, \cos\theta , \sin \theta)$\\
$a\frac{1}{\sqrt{2}}(0, -1, 1)$ & $a\frac{1}{{2}}(0, -1, \varphi)$ & $a(0, -\cos \theta, \sin \theta)$\\
$a\frac{1}{\sqrt{2}}(0, 1, -1)$ & $a\frac{1}{{2}}(0, 1, -\varphi)$ & $a(0, \cos \theta, -\sin \theta)$\\
$a\frac{1}{\sqrt{2}}(0, -1, -1)$ & $a\frac{1}{{2}}(0, -1, -\varphi)$ & $a(0, -\cos \theta, -\sin \theta)$\\
$a\frac{1}{\sqrt{2}}(1, 1, 0)$ & $a\frac{1}{{2}}(1, \varphi, 0)$ &$a(\cos \theta, \sin \theta, 0)$\\
$a\frac{1}{\sqrt{2}}(-1, 1, 0)$ & $a\frac{1}{{2}}(-1, \varphi, 0)$ & $a(-\cos \theta, \sin \theta, 0)$\\
$a\frac{1}{\sqrt{2}}(1, -1, 0)$ & $a\frac{1}{{2}}(1, -\varphi, 0)$ & $a(\cos \theta, -\sin \theta, 0)$\\
$a\frac{1}{\sqrt{2}}(-1, -1, 0)$ & $a\frac{1}{{2}}(-1, -\varphi, 0)$ & $a(-\cos \theta, -\sin \theta, 0)$\\
$a\frac{1}{\sqrt{2}}(1, 0, 1)$ & $a\frac{1}{{2}}(\varphi, 0, 1)$ & $a(\sin \theta, 0, \cos \theta)$\\
$a\frac{1}{\sqrt{2}}(1, 0, -1)$ & $a\frac{1}{{2}}(\varphi, 0, -1)$ & $a(\sin \theta, 0, -\cos \theta)$\\
$a\frac{1}{\sqrt{2}}(-1, 0, 1)$ & $a\frac{1}{{2}}(-\varphi, 0, 1)$ & $a(-\sin \theta, 0, \cos \theta)$\\
$a\frac{1}{\sqrt{2}}(-1, 0, -1)$ & $a\frac{1}{{2}}(-\varphi, 0, -1)$ & $a(-\sin \theta, 0, -\cos \theta)$\\
$a\frac{1}{\sqrt{2}}(0, 0, 0)$ & $a\frac{1}{{2}}(0, 0, 0)$ & $a(0, 0, 0)$\\
\hline\end{tabular}
\end{table}

\begin{figure}[htbp]

   \centerline{\hbox{
   \epsfxsize=5.0in
   \epsffile{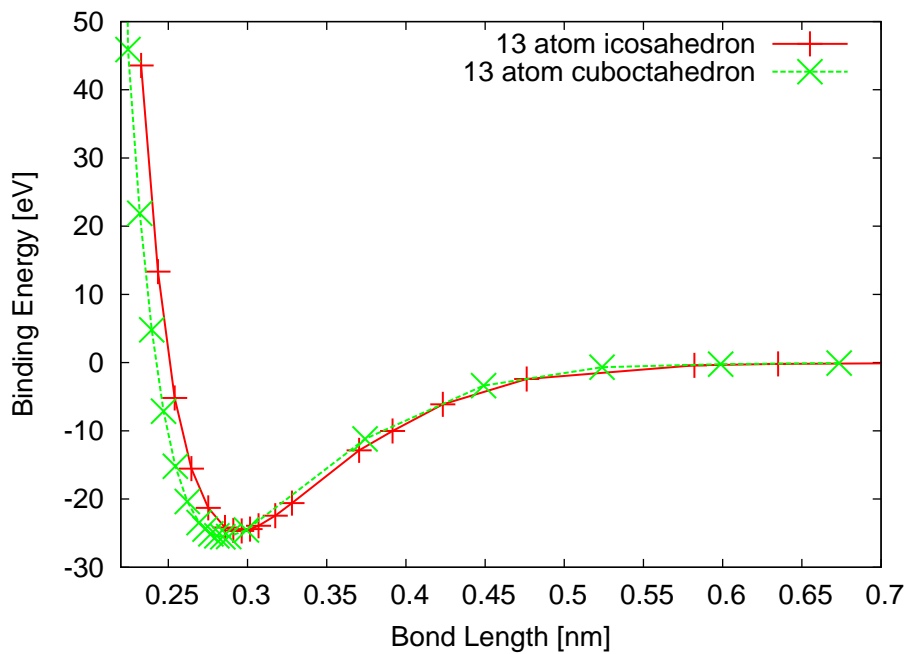}
     }
  }
 \caption{Binding energy as a function of bond length for 13 atom gold nanoclusters shaped like an icosahedron and cuboctahedron.
}

  \label{fig1}

\end{figure}

\begin{figure}[htbp]

   \centerline{\hbox{
   \epsfxsize=5.0in
   \epsffile{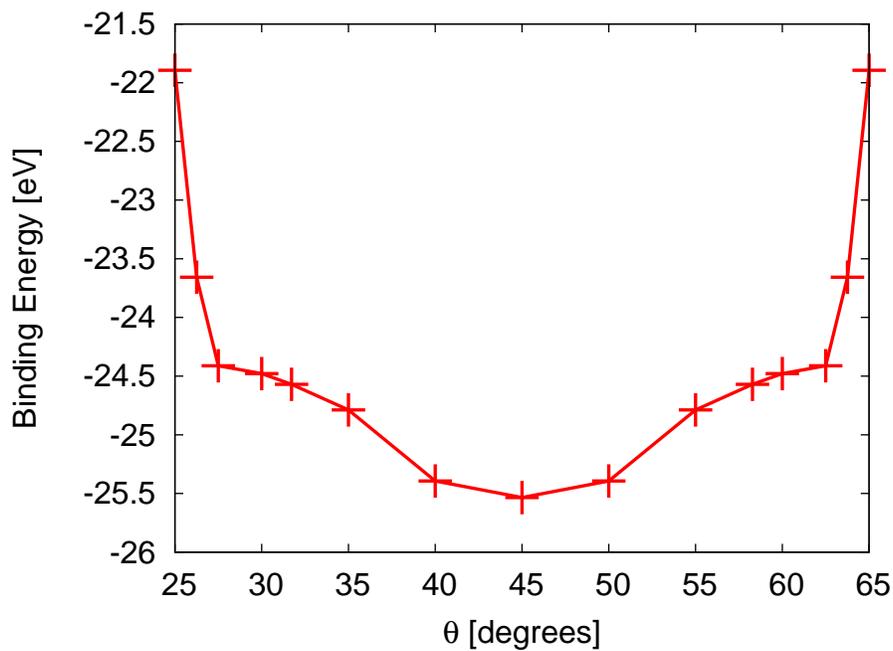}
     }
  }
 \caption{Binding energy as a function of the shape parameter $\theta$ for a 13 atom gold cluster. The minimum occurs at 45 degrees which corresponds to the shape of a cuboctahedron.
}

  \label{fig2}

\end{figure}

\begin{table}
\centering \caption{Data near the minimum of the binding potential for 13 atom gold nanoclusters.}
\label{pdtable3}
\begin{tabular}{|c|c|c|}
\hline
shape & bond length [nm]&  Binding Energy [eV] \\
\hline
cuboctahedron & 0.276897     &   -25.1064\\
cuboctohedron & 0.280639     &   -25.3946\\
cuboctohedron & 0.28438       &  -25.5349\\
cuboctohedron & 0.288122      &  -25.4893 \\
icosahedron  &	0.285756     &   -24.1683\\
icosahedron &	0.291047     &   -24.5989\\
icosahedron &	0.296339  &      -24.6706\\
icosahedron &	0.301631  &      -24.4062\\
\hline\end{tabular}
\end{table}

In this paper we primarily study $K = 2$ and $K =3$ which corresponds to 13 and 55 gold atoms respectively. Higher values of $K$ can also be studied and require large computational resources and scalable density functional codes. In section 3 we present some results for $K=4$ and 147 gold atoms.

Figure 2  shows the binding potential for the cuboctohedron and icosahedron structure. The difference in the potentials is rather small so we list the values near the bottom of the potential in Table 3. From Table 3 we see that the binding energy of the cuboctohedron shape for 13 gold atoms, -25.5349 eV, is greater in magnitude than that of the icosahedron, which was -24.6706 eV. Thus our calculations indicates that the cuboctahedron is more stable than the icosahedron for 13 gold atoms. The binding energy of the cuboctahedron shape minus the binding energy for the icosahedron shape for 13 gold atoms is
$\Delta E = -.8643$ eV.  This is in agreement with \cite{Haberlen} \cite{Wang}. Figure 3 shows the dependence of the energy on the shape parameter $\theta$. The potential shows a minimum at $\theta = 45$ degrees which corresponds to the cuboctohedron. Thus the cuboctahedron is stable with respect to shape transformations relative to the icosahedron. Indeed from the figure the icosahedron is not a local minimum and is unstabe with respect to shape perturbations for the 13 atom gold nanocluster.

\section{ Finite size and shape effects for 55 atom gold nanoclusters}

\begin{table}
\centering \caption{Coordinates for 55 gold nanoclusters with bond length $a$ and shape of cuboctahedron, icosahedron and intermediate structure with shape parameter $\theta$. Here $n$ is the number of atoms in each class of coordinates.}
\label{pdtable4}
\begin{tabular}{|c|c|c|c|}
\hline
cuboctohedron & icosahedron & intermediate & $n$\\
\hline
13 atom cuboctohedron & 13 atom icosahedron &  13 atom intermediate & 13\\
same as above scaled by 2 & same as above scaled by 2& same as above scaled by 2 & 12\\
$a\frac{1}{\sqrt{2}}(\pm 2, 0, 0)$ & $a\frac{1}{{2}}(\pm 2 \varphi, 0, 0)$ & $a( \pm 2 \sin \theta, 0, 0)$ & 2\\
$a\frac{1}{\sqrt{2}}(0, \pm 2, 0)$ & $a\frac{1}{{2}}(0, \pm 2 \varphi, 0)$ & $a(0, \pm 2 \sin \theta, 0)$ & 2\\
$a\frac{1}{\sqrt{2}}(0, 0, \pm 2)$ & $a\frac{1}{{2}}(0, 0, \pm 2 \varphi)$ &$a(0, 0, \pm 2 \sin \theta)$ & 2\\
$a\frac{1}{\sqrt{2}}(\pm 2, \pm 1, \pm 1)$ & $a\frac{1}{{2}}(\pm (1 + \varphi), \pm \varphi, 1)$ & $a(\pm (\cos \theta + \sin \theta),  \pm \sin \theta, \pm \cos \theta)$ & 8\\
$a\frac{1}{\sqrt{2}}(\pm 1, \pm 2 , \pm 1)$ & $a\frac{1}{{2}}(\pm 1, \pm (1 + \varphi), \pm \varphi)$ & $a(\pm \cos \theta,\pm(\cos \theta + \sin \theta), \pm \sin \theta)$ & 8\\
$a\frac{1}{\sqrt{2}}(\pm 1, \pm 1, \pm 2)$ & $a\frac{1}{{2}}(\pm \varphi , \pm 1, \pm (1 + \varphi))$ & $a( \pm \sin \theta, \pm \cos \theta, \pm ( \cos \theta + \sin \theta))$ & 8\\
\hline\end{tabular}
\end{table}

One can proceed in a similar manner for the 55 atom gold cluster. The cuboctahedron and icosahedron 55 atom gold nanocluster are shown in Figure 4 and coordinates are given in Table 4. The binding energy a function of bond length is shown in Figure 5 and Table 5  for the cuboctahedron and icosahedron respectively. The data indicates that the icosahedron is lower in energy than the cuboctahedron with $\Delta E = E_{cuboctahedron} - E_{icosahedron} = 1.28019$ eV. 

\begin{figure}[tbp]
   \centerline{\hbox{
   \epsfxsize=3.0in
   \epsffile{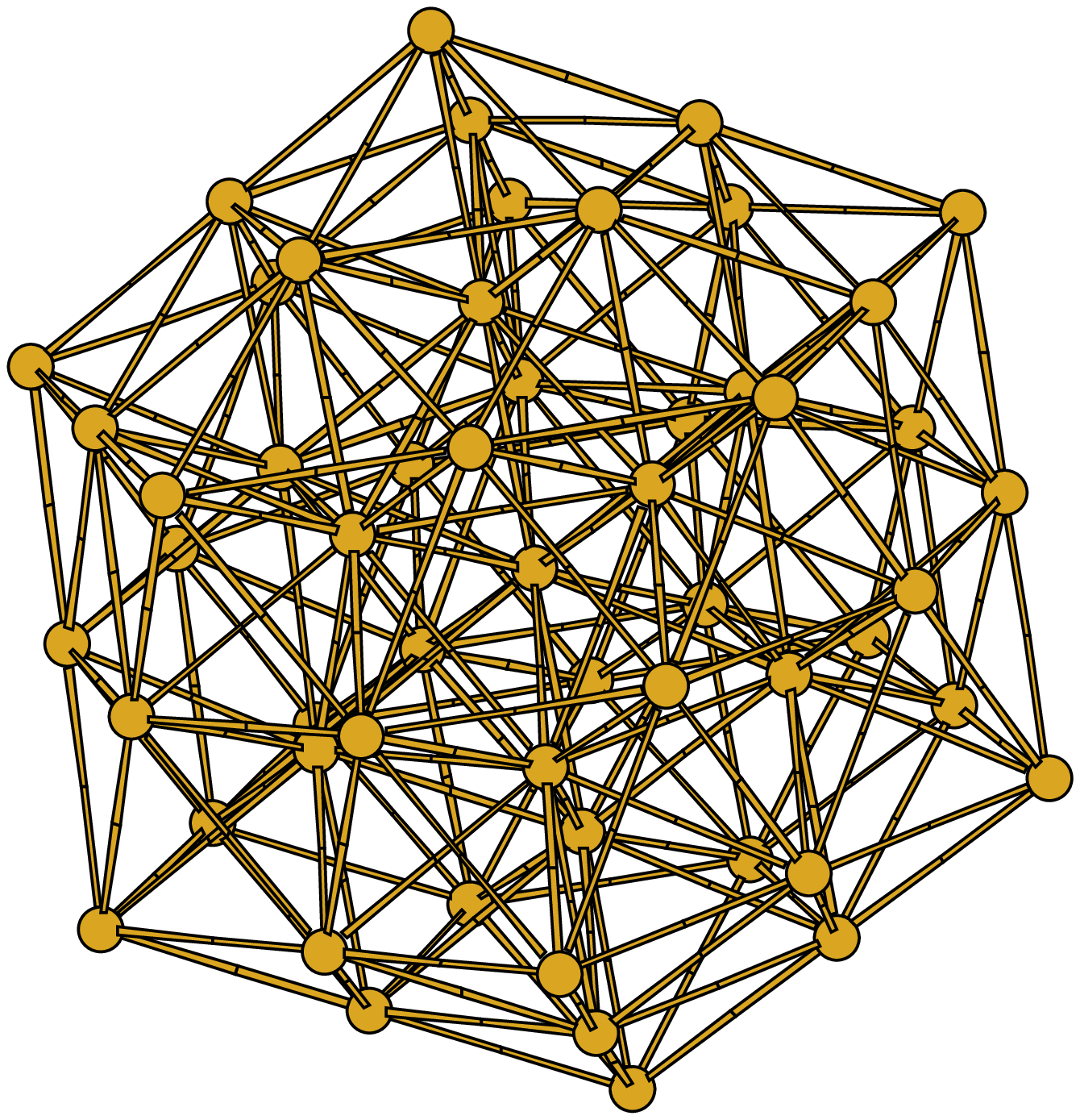}
   \epsfxsize=3.0in
   \epsffile{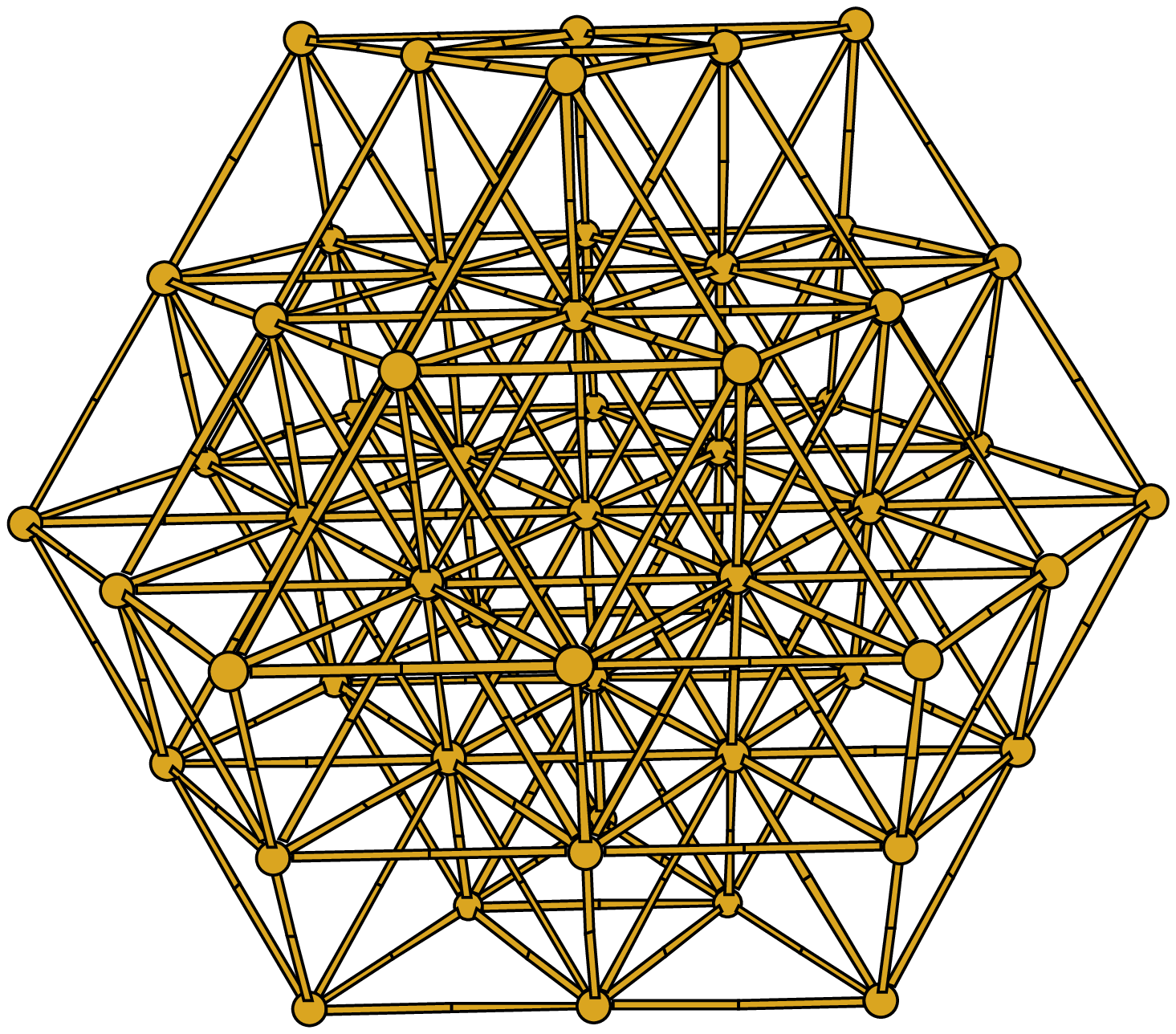}
     }
  }
 \caption{55 gold atom nanocluster in the shape of an icosahedron (left) and cuboctahedron (right).}

  \label{fig3}

\end{figure}

\begin{figure}[htbp]

   \centerline{\hbox{
   \epsfxsize=5.0in
   \epsffile{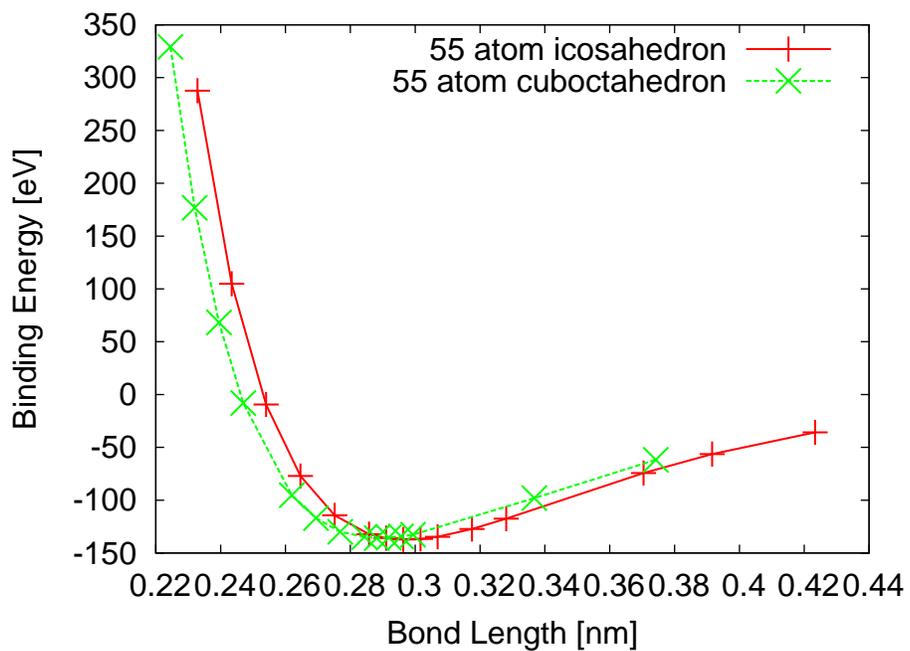}
     }
  }
 \caption{Binding energy as a function of the bond length for a 55 atom gold cluster.}

  \label{fig4}

\end{figure}

\begin{figure}[htbp]

   \centerline{\hbox{
   \epsfxsize=4.0in
   \epsffile{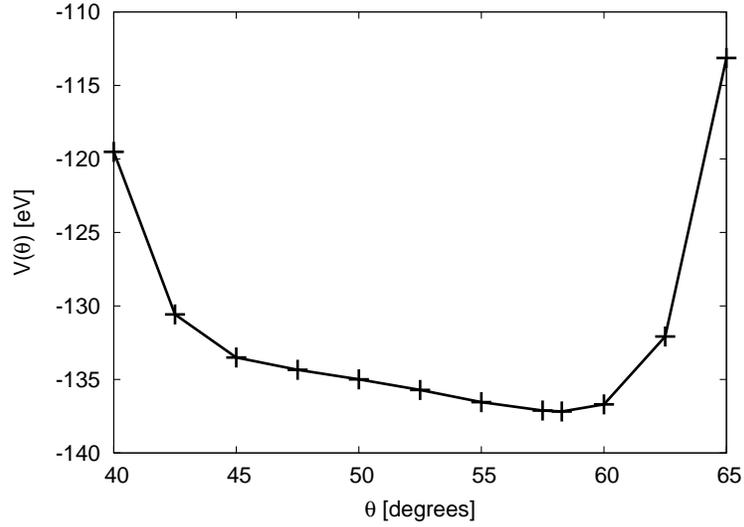}
     }
  }
 \caption{Binding energy as a function of the shape parameter $\theta$ for a 55 atom gold cluster. The minimum occurs at $58.282525589$ degrees which corresponds to the shape of a icosahedron.
}

  \label{fig5}

\end{figure}

\begin{table}
\centering \caption{Data near the minimum of the binding potential for 55 atom gold nanoclusters.}
\label{pdtable5}
\begin{tabular}{|c|c|c|}
\hline
shape & bond length [nm]&  Binding Energy [eV] \\
\hline
cuboctahedron &   0.276897  &   -129.961\\
cuboctohedron &  0.288122   &   -135.901\\
cuboctohedron &   0.291864  &  -135.319\\
cuboctohedron &   0.299348   &  -132.635 \\
icosahedron  &	0.285756     &  -132.143  \\
icosahedron &	 0.291047   &   -135.901 \\
icosahedron &	 0.296339 &      -137.181\\
icosahedron &	  0.301631&      -136.643\\
icosahedron &	   0.306923  &   -134.659\\
\hline\end{tabular}
\end{table}

\begin{table}
\centering \caption{Data near the minimum of the binding potential for 147 atom gold nanoclusters.}
\label{pdtable6}
\begin{tabular}{|c|c|c|}
\hline
shape & bond length [nm]&  Binding Energy [eV] \\
\hline
cuboctahedron &   0.28438 &  -383.226 \\
cuboctohedron &    0.288122 &  -387.14  \\
cuboctohedron &  0.291864  & -387.988 \\
cuboctohedron &    0.295606  &  -386.206 \\
icosahedron  &	  0.291047  &  -385.21  \\
icosahedron &	   0.296339 &   -391.058 \\
icosahedron &	0.301631  &     -391.033 \\
\hline\end{tabular}
\end{table}

In order to further examine the stability of 55 atom gold clusters with respect to change in shape we plot the binding energy as a function of the shape parameter $\theta$ in figure 8. The angle $\theta$ is defined similar to the 13 atom case as being 45 degrees for the cuboctohedron, $\arctan \varphi$ for the icosahedron and intermediate values in between. A simple prescription to define the shape parameter is to replace the occurrence of the golden ratio $\varphi$ in the list of icosahedral coordinates with $\tan \theta$ and scale all the 55 coordinates proportional to $\cos \theta$ to construct the interpolating nanostructures. The same prescription works also for the 13 atom case as indicated in tables 2-4. The inner bond length for the 55 atom case is fixed at $0.951057 a_0$ where $a_0=0.296339$ nm. The resulting plot indicates the the icosahedral structure is at lower energy than the cuboctahedral structure for nanoclusters consisting of 55 gold atoms with respect to change of shape.

It should be emphasized however that we have just investigated the relative stability between two well known structures, the cuboctohedron and the icosahedron. There are many more structures that do not fit into these categories or even the interpolation between them. Some of these have lower energy \cite{Apra}. Nevertheless the ability to computationally determine the potentials for size and shape of nanoclusters gives a first simplified  look at the so called energy landscape of theses models. The demonstration of the ability to use efficient density functional codes and parallel computing to map out this landscape is one of the main results of this paper.

One can extend the results of this paper to larger gold clusters. Table 6 shows the results of a binding energy calculations of  147 atom gold clusters. The table indicates that for the 147 gold atom case the icosahedron is lower in energy than the cuboctahedron with $\Delta E = E_{cuboctahedron} - E_{icosahedron} = 3.07$ eV. 

\section{Calculation of $Au_{55}O_2$ binding energy}
\begin{figure}[htbp]

   \centerline{\hbox{
   \epsfxsize=4.0in
   \epsffile{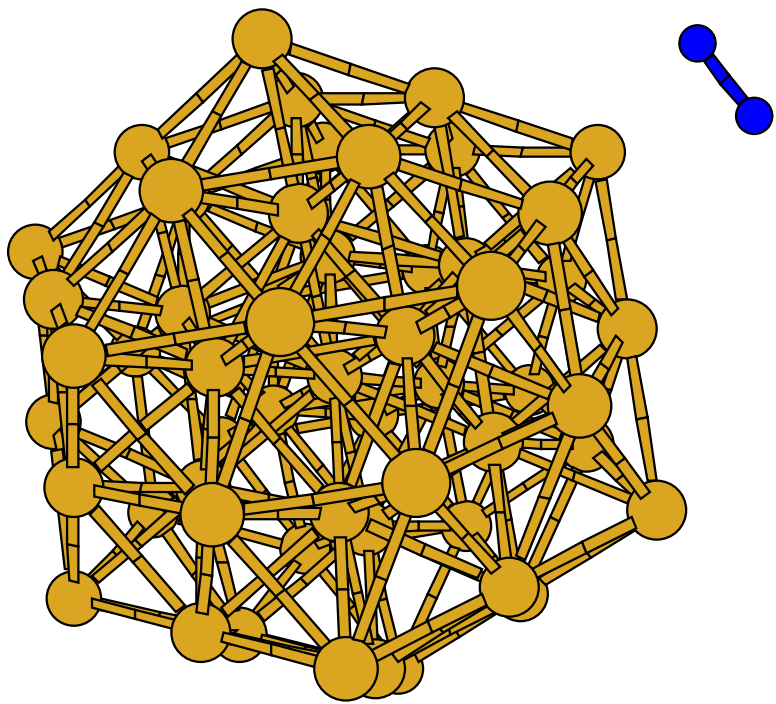}
     }
  }
 \caption{A 55 atom gold icosahedron nanoparticle interacting with $O_2$.}

  \label{fig6}

\end{figure}

\begin{figure}[htbp]

   \centerline{\hbox{
   \epsfxsize=4.0in
   \epsffile{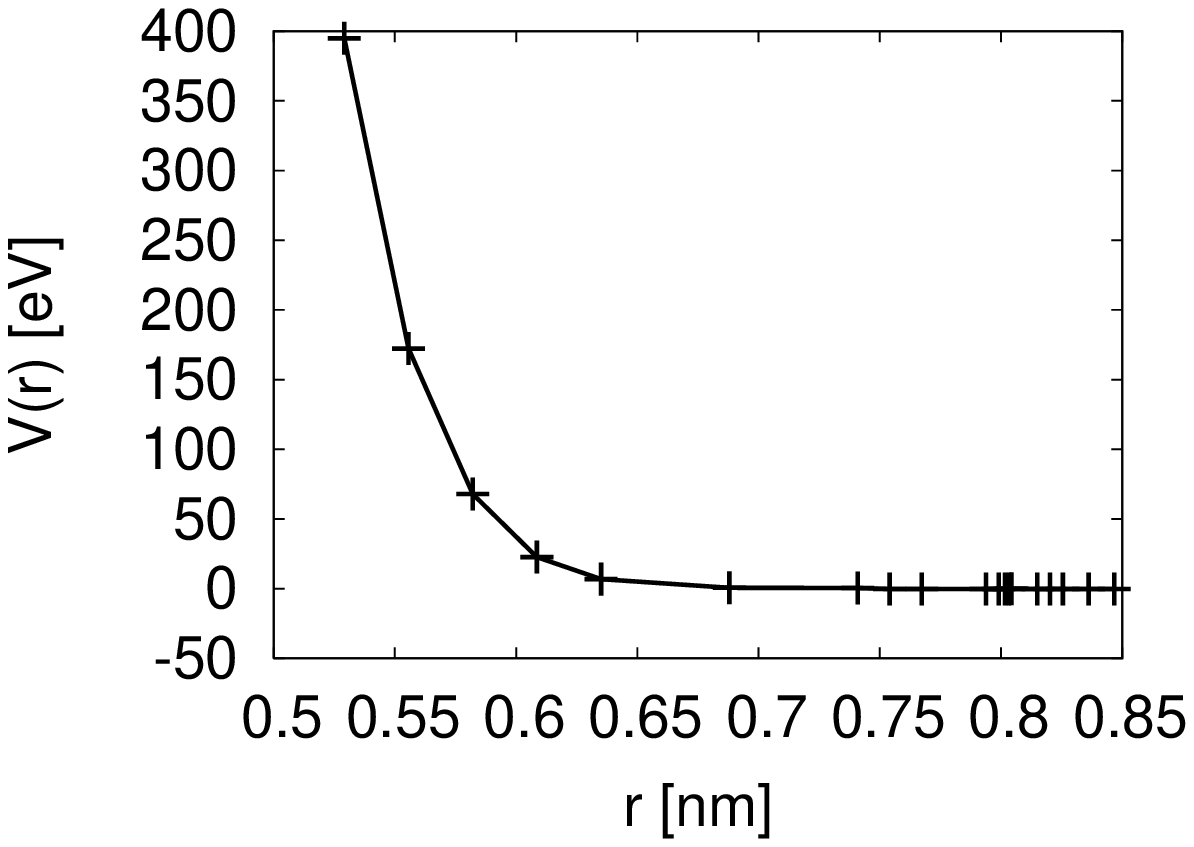}
     }
  }
 \caption{Binding energy as a function of the distance to the center of mass of the $O_2$ molecule to the center of mass of the $Au_{55}$ nanoparticle along the x axis between .5 and .85 nm.}

  \label{fig7}

\end{figure}

\begin{figure}[htbp]

   \centerline{\hbox{
   \epsfxsize=4.0in
   \epsffile{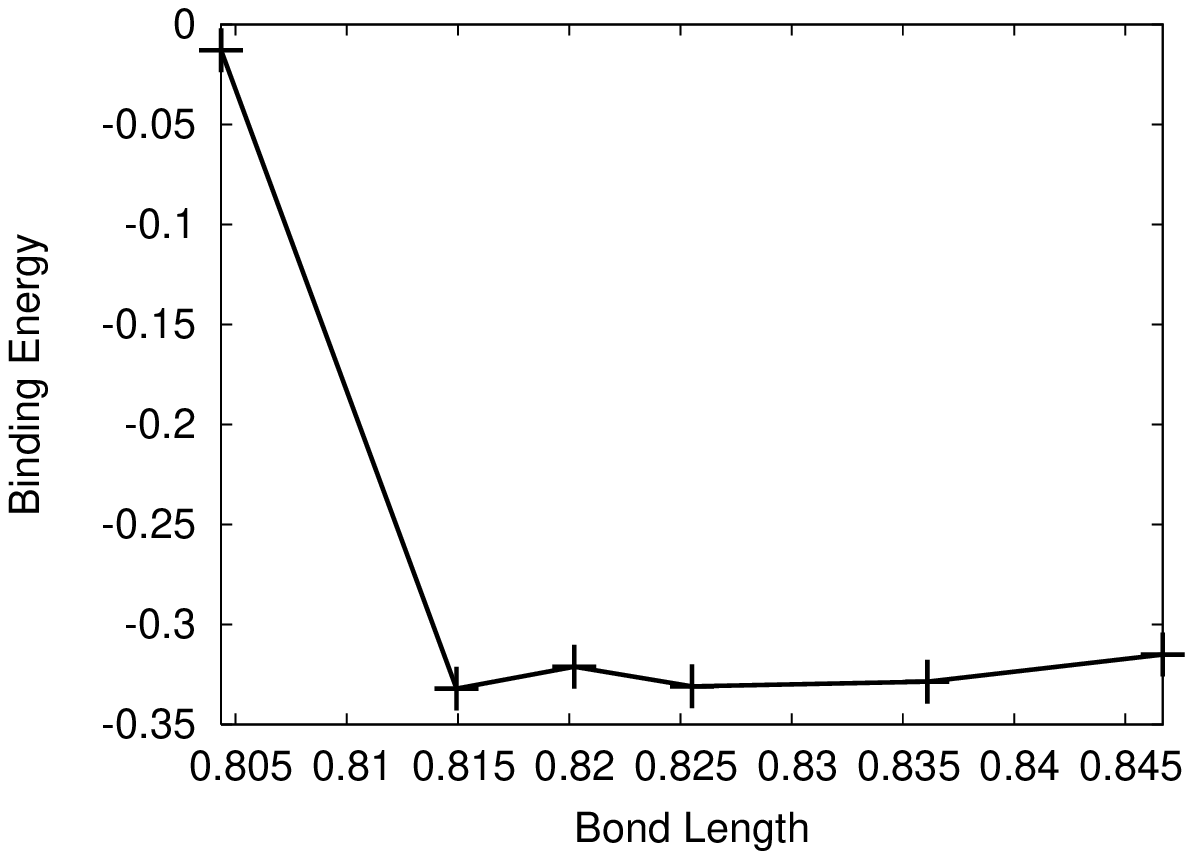}
     }
  }
 \caption{Binding energy as a function of the distance to the center of mass of the $O_2$ molecule to the center of mass of the $Au_{55}$ nanoparticle along the x axis between .805 and .845 nm.}

  \label{fig8}

\end{figure}

\begin{figure}[htbp]

   \centerline{\hbox{
   \epsfxsize=4.0in
   \epsffile{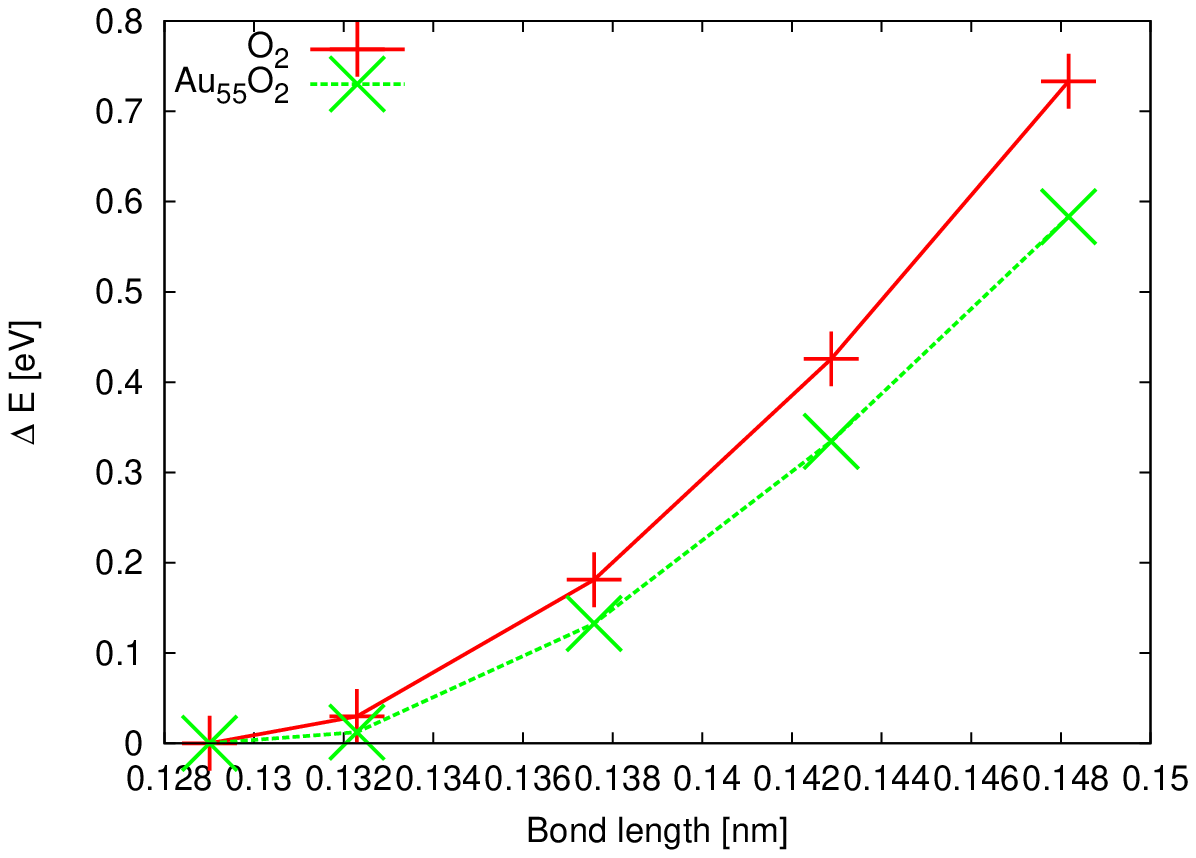}
     }
  }
 \caption{Difference in energy from the lowest energy state as a function of the distance between the two oxygen atoms.}
  \label{fig9}

\end{figure}

Besides pure gold clusters another area of interest is the use of gold clusters as oxygen catalysts involving the disassociation of $O_2$ molecules. 

In  \cite {Boyen} X-ray spectroscopy was used to show that 55 atom gold clusters have a maximum oxidation resistance and suggests that they may be effective oxidation catalysts. In \cite{Okumura} hybrid density functional calculations were done which suggests that the surfaces of small gold clusters are active cites for catalytic interactions. Among various systems they studied $Au_{13}O_2$ clusters. In \cite{Turner} electronic structure of 55 atom gold nanoparticles and catalytic activity from the modified electronic structure for small nanoclusters ($\sim 1.4$ nm). In \cite{Roldan} density functional calculations of $Au_{55}$ clusters were shown not to dissociate $O_2$

To study these effects we calculated the binding energy of the $Au_{55}O_2$ composite system using NWChem. The $Au_{55}O_2$ molecule is shown in Figure 7. The results for the binding potential are shown in Figure 8  and 9. These figures plot the energy as as a function of the distance of the center of mass of the $O_2$ along the $x$ axis using the coordinates of $Au_{55}$ from Table 4. Figure 8 shows a very flat potential away from .8 nm so we show an expanded version of the potential in figure 9 from .805 to .845 nm. This shows a minimum at .815  nm and a weak binding energy of -0.332135 eV.

More interesting from the point of catalysis is to fix the center of mass distance of the $O_2$, vary the distance between the two oxygens and study the energy as a function of this bond length. This is shown in Figure 10 together with the the energy as function of bond length for free standing $O_2$ not in the presence of the gold nanoparticle. Here we fixed the distance from the origin  to the center of mass of the $O_2$ to be .815 nm. For all the data points that we studied in Figure 10 the presence of the $Au_{55}$ nanoparticle lowered the energy and made the potential flatter as a function of separation of the two oxygens then would have been the case without the nanoparticle. This can be seen to give support to the idea that gold nanoparticle can be used to assist the breaking of $O_2$, but from the figure the difference is small , less that .1 eV, so that more detailed studies are necessary to be definitive. One can also study the effect of different orientation of the $O_2$  with respect to the gold nanoparticle as this can effect the energy of the system as well as the binding of the $O_2$, rather than bring in the $O_2$ along the x axis as we have done.

\section{Conclusion}

In this paper we have begun a study of the energy landscape for gold nanoclusters using density functional theory run on large parallel computers. By restricting ourselves to cuboctohedral and icosahedral geometries as well as intermediate geometries we have explored a small subspace of low energy gold nanostructures. We found that the cuboctahedral shape was lower energy than the icosahedral shape for 13 atoms and the opposite was the case for 55 and 147 atoms. We also estimated the energy difference. More significantly we were able to calculate the potential energy as a function of bond length and shape by introducing a shape parameter $\theta$. Finally we were able to calculate the effect of the presence of nanogold on the binding energy of $O_2$ showing that it leads to weaker binding which may have applications for the use of nanogold for catalysis. In the future it will be interesting to expand the configuration space of geometries that can be efficiently probed for gold nanoclusters. Because the chemistry and material properties of these clusters depend strongly on geometry this is likely to have practical applications. Rapidly expanding computational resources dedicated to nanoscience indicates that this will be possible in the near future.

\section*{Acknowledgments}
This manuscript has been authored in part by Brookhaven Science Associates, LLC, under Contract No. DE-AC02-98CH10886 with the U.S. Department of Energy.

\end{document}